\documentclass{article}
\usepackage[preprint]{spconf}
\usepackage{amsmath,graphicx, subcaption, hyperref}

 \copyrightnotice{\copyright\ IEEE 2023}
\toappear{To appear in {\it Proc.\ ICASSP 2023,
                    June 04-10, 2023, Rhodes Island, Greece}}

\title{SELF-SUPERVISED REPRESENTATIONS FOR SINGING VOICE CONVERSION}
\name{\parbox{\linewidth}{\centering Tejas Jayashankar$^{\,1,*}$ \thanks{* Work performed during internship at Meta AI}, Jilong Wu$^{\,2}$, Leda Sari$^{\,2}$, David Kant$^{\,2}$,\\
{Vimal Manohar}$^{\,2}$, {Qing He}$^{\,2}$}}

\address{
$^1\,$Massachusetts Institute of Technology\quad
$^2\,$Meta AI
}
\begin{document}
\ninept

\maketitle

\begin{abstract}
A singing voice conversion model converts a song in the voice of an arbitrary source singer to the voice of a target singer.  Recently, methods that leverage self-supervised audio representations such as HuBERT and Wav2Vec 2.0 have helped further the state-of-the-art.  Though these methods produce more natural and melodic singing outputs, they often rely on confusion and disentanglement losses to render the self-supervised representations speaker and pitch-invariant.  In this paper, we circumvent disentanglement training and propose a new model that leverages ASR fine-tuned self-supervised representations as inputs to a HiFi-GAN neural vocoder for singing voice conversion.  We experiment with different $f_0$ encoding schemes and show that an $f_0$ harmonic generation module that uses a parallel bank of transposed convolutions (PBTC) alongside ASR fine-tuned Wav2Vec 2.0 features results in the best singing voice conversion quality.  Additionally, the model is capable of making a spoken voice sing.  We also show that a simple $f_0$ shifting scheme during inference helps retain singer identity and bolsters the performance of our singing voice conversion model. Our results are backed up by extensive MOS studies that compare different ablations and baselines.
\end{abstract}
\section{Introduction}
\label{sec:intro}

The goal of a singing voice conversion task is to convert a song in the voice of a source singer, say A, to the voice of a target singer, say B.  The conversion model should retain the singer-invariant content of the song such as the linguistic content and only change the singer dependent characteristics such as the $f_0$ and the singer's timbre.

Early work on singing voice conversion used HMM based architectures \cite{saino2006hmm, oura2010recent, nakamura2014hmm} for modeling the latent linguistic and melodic content in singing voices.  Subsequent work based on parametric statistical models tackled the problem of parallel singing voice conversion, wherein paired singing samples in the voice of both the source singer and target singer are available \cite{kobayashi2014statistical, kobayashi2015statistical}.  Learning parallel singing voice conversion models is difficult as it is expensive to collect parallel samples from multiple different singers.   

To circumvent the data collection difficulties associated with parallel singing voice conversion models, recent work has focused on non-parallel singing voice conversion. In addition to this, the success of neural network architectures, specifically neural vocoders such as WaveNet, \cite{oord2016wavenet}, WaveRNN \cite{kalchbrenner2018efficient} and HiFi-GAN \cite{kong2020hifi}, has once again made the research area of voice conversion a hot topic. The method of Unsupervised Singing Voice Conversion \cite{nachmani2019unsupervised} employs an autoencoder architecture with a WaveNet decoder to convert between a fixed set of singers.  The model is trained with a domain confusion loss to extract singer-invariant features from the encoder.  The decoder is conditioned on the target singer identity to synthesize a song in the desired voice.  PitchNet \cite{deng2020pitchnet} builds on this model by employing an additional adversarial pitch confusion term to extract pitch-invariant and singer-invariant features from the encoder.  The WaveNet decoder is conditioned on additional source $f_0$ information which can help improve the quality of the converted singing voice and retain important melodic information.

Rather than relying on a confusion loss, the method of Unsupervised Cross-Domain Singing Voice Conversion \cite{polyak2020unsupervised} trains a GAN based architecture with a WaveNet generator and a convolutional discriminator.  The generator synthesizes singing audio given $f_0$ information, a speaker embedding and intermediate ASR features, which are assumed to be disentangled.  During inference the source speaker embedding can be swapped out with a target speaker embedding to perform voice conversion.

The use of phonetic posteriograms (PPGs), namely the penultimate layer output of as ASR model, has gained large traction in the voice conversion community. The model in \cite{wang2021towards} uses PPG outputs to model the linguistic content of the source singing sample and the state-of-the-art HiFi-GAN vocoder for synthesis.  They build upon \cite{polyak2020unsupervised} by using HuBERT \cite{hsu2021hubert} features to model the melodic content.  These self-supervised features are learned embeddings that encapsulate linguistic and melodic information within windows of the input audio. They are used as additional melodic conditioning to the model on top of the source $f_0$ for improving the singing voice quality. The HuBERT features are fine-tuned to be pitch-invariant and speaker-invariant to avoid mismatch between the input $f_0$ and the pitch encoded within HuBERT features.

In this paper we further investigate the benefits of pre-trained self-supervised features for singing voice conversion/synthesis.  Our main contribution is a new singing voice conversion model that uses a parallel bank of transposed convolutions (PBTC) for encoding $f_0$ and ASR fine-tuned self-supervised features for modeling the singer invariant features.  Furthermore, to retain target singer similarity, we revisit a simple $f_0$ shifting technique from voice conversion literature and demonstrate how it can be used with our model to boost target singer identity similarity after conversion.  We perform extensive MOS studies to evaluate different architectural designs for singing voice conversion.

Our paper is organized into five sections.  In Section 2, we explain our proposed model architecture and introduce concepts related to $f_0$ encoding, self-supervised representations and neural vocoders.  In Section 3, we provide implementation details and report results from the MOS studies we performed to evaluate our models.  Finally, we wrap up with a conclusion, some remarks on future work and acknowledgements.


\begin{figure*}[!thb]
    \centering
    \includegraphics[scale=0.30]{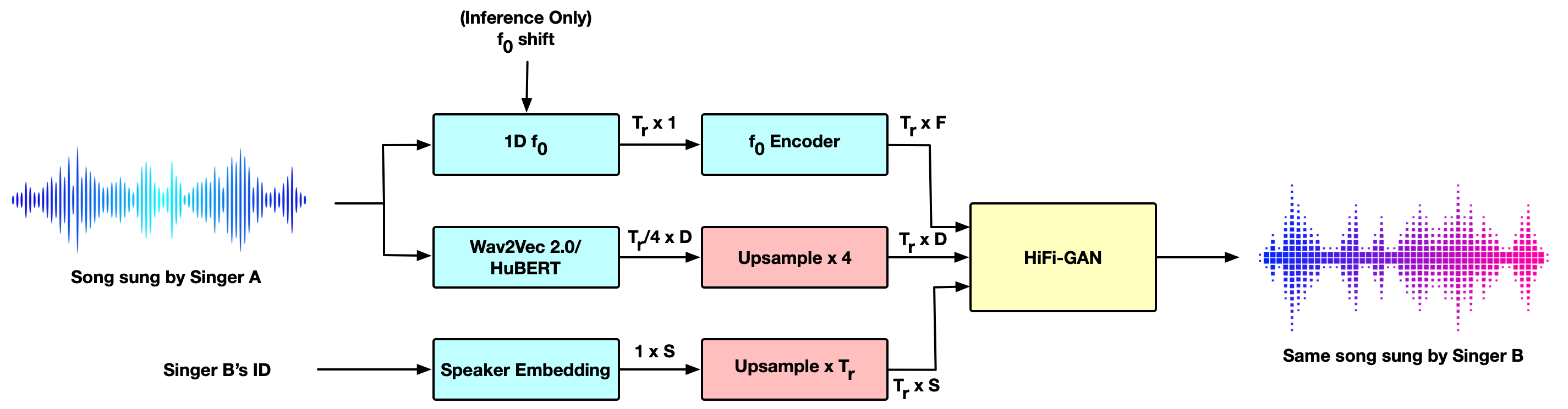}
    \caption{The overall architecture of our proposed singing voice conversion system.  The HiFi-GAN decodes a new singing sample in the voice of Singer B given a singer embedding correponding to this singer and content (pitch + self-supervised) features extracted from the original song in Singer A's voice.  Note that the features are extracted at different frequencies and need to be upsampled appropriately before concatenation.  During inference, $f_0$ scale and shift is performed to match the source $f_0$ with the target singer's $f_0$ distribution.\vspace{-1em}}
    \label{fig:architecture}
\end{figure*}

\section{Method}
\label{sec:method}

The overall architecture of our proposed model is shown in Figure \ref{fig:architecture}.  The model is composed of an $f_0$ feature module, a singer embedding module, a self-supervised feature extractor and a HiFi-GAN decoder.  In this section we provide more details about each of these components and the overall working of the system.

\subsection{System Overview}
\label{sec:system overview}

Our model operates by extracting two different types of features from the source audio: 1) the source $f_0$ and 2) self-supervised feature embeddings of dimension $D$.  The $f_0$ is further processed by embedding it into a continuous latent space of dimension $F$ using an $f_0$ feature encoder. Additionally, a target singer ID is provided to the model to indicate in whose voice the song must be sung.  The singer ID is fed to an embedding model, which in practice is a learned look up table (LUT) of size $N \times S$ where the dimension of each embedding is $S$ and $N$ is the number of singers. The three different features are then upsampled and concatenated together resulting in a feature with dimensions $T_r \times (F + D + S)$ where $T_r$ is the total number of frames in the input source audio.  The HiFi-GAN decoder then synthesizes a singing voice at 24 kHz from the concatenated features.  

During training, the singer ID is unchanged and the task is to decode the source audio from the concatenated features.  During inference, we perform voice conversion by swapping the source singer ID with a different target singer ID.  We additionally shift the source $f_0$ to the target $f_0$'s domain by performing a simple scale and shift operation (see Section \ref{sec:$f_0$ shifting}).

\subsection{Self-supervised Features: Wav2Vec 2.0 and HuBERT}
\label{sec:self supervised features}

Our model uses self-supervised features to encode the speaker-invariant characteristics of a singing voice such as the linguistic/speech content and pitch independent melodic content (e.g., timing information such as the duration of phonemes).

Taking inspiration from recent voice conversion models \cite{polyak2021speech}, we experiment with the recent and popular Wav2Vec 2.0 \cite{baevski2020wav2vec} and HuBERT \cite{hsu2021hubert} embeddings.  Both Wav2Vec 2.0 and HuBERT consist of a CNN feature encoder that encodes the input audio frames and a BERT-like \cite{devlin2018bert} transformer trained with a masked predictive task that enforces that the learned encodings contain strong acoustic and linguistic content given the past and future context.

Wav2Vec 2.0 is trained with a contrastive loss to correctly predict the quantized CNN encoder output at time step $t$ from a set of distractors.  On the other hand, HuBERT training forces the model to correctly predict the quantized latent from a random layer $k$ of the transformer at time step $t$ amongst a set of distractors taken at different time steps from the same layer --- hence the name \textbf{H}idden \textbf{u}nit BERT.  In both settings, the continuous input at time step $t$ to the transformer is masked out.

We also experiment with ASR fine-tuned Wav2Vec 2.0 features.  The pre-trained Wav2Vec 2.0 features are fine-tuned on an ASR prediction task by fusing a softmax layer on top of the BERT transformer with a CTC \cite{graves2006connectionist} loss.  Some recent voice conversion models have relied on ASR outputs such as phonetic posteriograms (PPGs) for modeling the linguistic content of singing voices.  We postulate that the ASR fine-tuning task helps preserve linguistic content in the embeddings akin to PPGs.  These embeddings also contains important singer-invariant melody information that would otherwise need to be extracted using a special encoder or via disentanglement training/confusion losses as in \cite{nachmani2019unsupervised, deng2020pitchnet, wang2021towards}.

\subsection{$f_0$ Feature Encoder}
\label{sec:$f_0$ feature encoder}

We initially experimented with 1D $f_0$ features, i.e., $F=1$.  However, the HiFi-GAN synthesized singing outputs sounded monotonic and off-pitch.  In most cases the desired prosody was not synthesized.  Subsequently, we experimented with two $f_0$ feature encoders to learn higher dimensional $f_0$ representations as described next. 
\vspace{-0.5em}
\subsubsection{Q-LUT Embedding Module}
\label{sec:q-lut}

The continuous 1D $f_0$ sequence is first mapped to the range $[0, 1]$ using mean and variance normalization.  The entries are then uniformly quantized to $L$ bins.  A look up table (LUT) with $L$ entries of dimension $F$ is learned end-to-end with the HiFi-GAN decoder to learn a rich pitch embedding for each possible quantized value of the 1D $f_0$.  An illustration of the architecture is shown in Figure \ref{fig:qlut}.
\vspace{-0.5em}
\subsubsection{PBTC Harmonic Generation Module}
\label{sec:pbtc}

\begin{figure*}[t!]
    \centering
    \begin{subfigure}[t]{0.4\textwidth}
        \centering
        \includegraphics[scale=0.135]{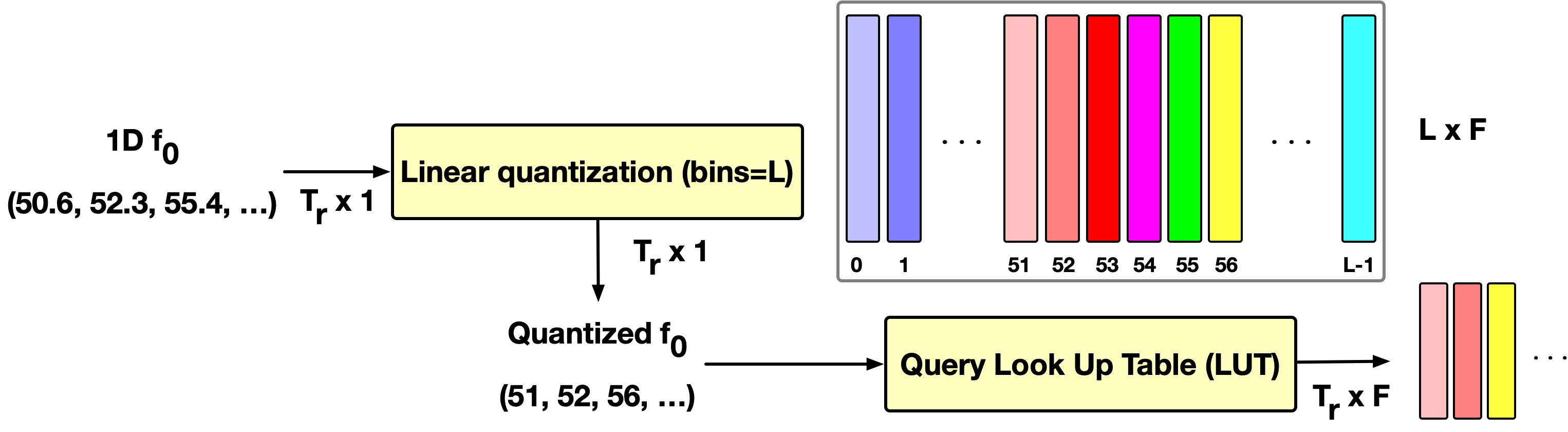}
        \caption{Q-LUT}
        \label{fig:qlut}
    \end{subfigure}%
    ~
    \begin{subfigure}[t]{0.6\textwidth}
        \centering
        \includegraphics[scale=0.135]{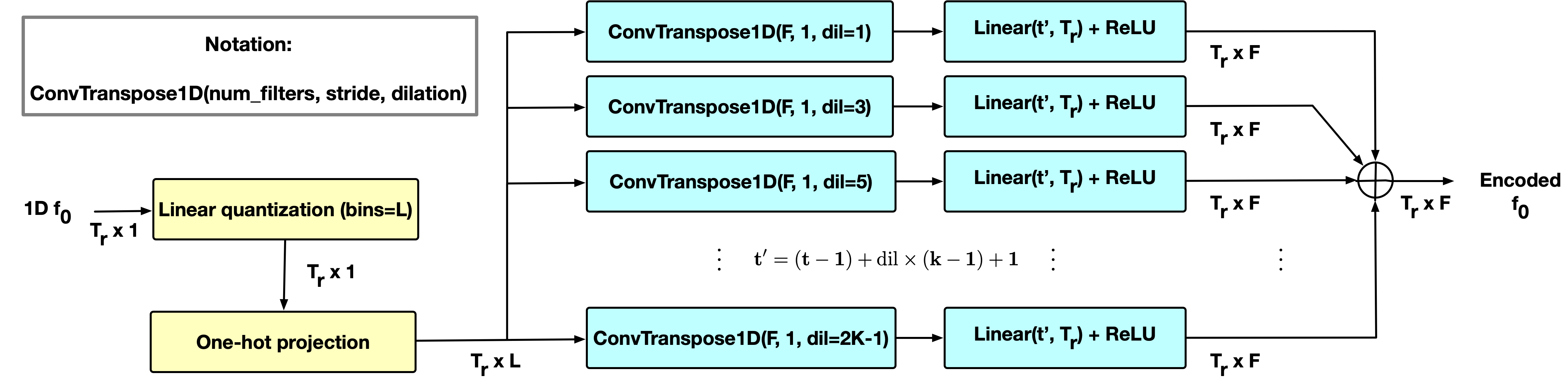}
        \caption{PBTC}
        \label{fig:pbtc}
    \end{subfigure}
    \caption{Architectures of the Q-LUT and PBTC $f_0$ encoding models that we experiment with in our system.}
\vspace{-1.5em}
\label{fig:f0 encoders}
\end{figure*}

The parallel bank of transposed convolutions (PBTC) was introduced in \cite{webber2020hider} for generating harmonic information from 1D $f_0$ sequences.  The continuous 1D $f_0$ sequence is again globally normalized and quantized to $L$ bins.  The quantized sequence of length $T_r$ is then one-hot projected to an embedding of shape $T_r \times L$.  This embedding is parallely processed with $K$ different 1D transposed convolutions with linearly increasing dilation rates and $F$ filters.  Since the input embedding is sparse (due to the one-hot encoding), the transposed convolutions learn appropriate filters to apply at the non-zero positions to generate rich harmonic information at different rates.  The output of each transposed convolution is linearly projected to the same duration $T_r$ and then summed together resulting in a new feature of shape $T_r \times F$.  An illustration of the architecture is shown in Figure \ref{fig:pbtc}.  As will be described in Section \ref{sec: results}, MOS studies indicate that the PBTC architecture results in higher quality singing audio in comparison to the Q-LUT method.
\vspace{-1em}

\subsection{Synthesis with the HiFi-GAN Vocoder}
\label{sec:hifigan}

We use a slightly modified version of the original HiFi-GAN \cite{kong2020hifi} for synthesizing the singing voice from the input features.  All features are upsampled to the same rate before being fed to the generator. 

In our experiments the inputs to the generator are upsampled to a rate of 200 Hz.  The generator $G$ consists of four upsampling blocks with upsample rates of [3, 4, 5, 8].  Each upsampling block consists of three residual blocks with kernel sizes of [3, 7, 11] and dilation rates of [1, 3, 5] per residual block.  The discriminator comprises two networks: a Multi-Period Discriminator (MPD) and a Multi-Scale Discriminator (MSD).  The MPD consists of five sub-discriminators each operating on equally spaced samples.  The five sub-discriminators use different spacings of [2, 3, 5, 7, 11]. The MSD uses three sub-discriminators operating at three different scales of the input signal: original scale, $\times 2$ downsampled signal and $\times 4$ downsampled signal.
\vspace{-1em}

\subsection{$f_0$ Shifting During Inference}
\label{sec:$f_0$ shifting}

During training of our model, the speaker embedding LUT is learned end-to-end with the $f_0$ encoder and the HiFi-GAN vocoder, and qualitative experiments suggested that the learned singer embedding internally encodes the moments of the corresponding singer's $f_0$ distribution and hence are not disentangled. However, during inference, the source singer embedding can only be swapped out for a target singer embedding under the assumption that the embedding is disentangled from the $f_0$.  Attempts at disentangling the $f_0$ and singer embedding features led to poor results.  Instead, a simpler method that aligns the $f_0$ from the source singer with the internal $f_0$ information encoded in the target singer embedding worked well.

Assume that the $f_0$ distribution for singer A is a normal distribution $f_0^A \sim \mathcal{N}(\mu_A, \sigma_A)$ and that the $f_0$ distribution for singer B is $f_0^B \sim \mathcal{N}(\mu_B, \sigma_B)$.  Then given a source $f_0$ sequence, say $f_A$ from singer A, we shift it to match the $f_0$ distribution of singer B during inference.
\begin{align}
    f^{shift}_A = \frac{\sigma_B}{\sigma_A}(f_A - \mu_A) + \mu_B.
\end{align}

\section{Experiments}
\label{sec:experiments}

This section details the dataset, implementation details and results from different Mean Opinion Scores (MOS) studies that ran.\vspace{-0.5em}

\subsection{Dataset}
\label{sec:dataset}

We trained our model on a dataset consisting of speech and singing samples. We found that training only on singing samples resulted in poor synthesized audio quality.  We used an internal speech dataset consisting of $\sim$200 hours of English clean speech recorded in a voice production studio by contracted professional voice talents. Our singing voice dataset consists of 10+ hours of samples taken from the NUS-48E dataset \cite{duan2013nus}, the Children's Song Dataset (CSD) \cite{choi2020children} and the AmericanSong corpus licensed from SpeechOcean \footnote{https://en.speechocean.com/datacenter/details/3069.html}.  We use a 70-30 train-test split for training and evaluating our models.  All speech and singing samples are sampled at 24 kHz.  Our dataset consists of total of 17 different singers and over 20 different speakers.  \vspace{-0.5em}
\vspace{-1em}
\subsection{Implementation Details}
\label{sec:implementation details}

We use the CREPE \cite{kim2018crepe} pitch tracker to extract 1D $f_0$. The extraction range is set to [50, 800] Hz, where the upper bound was chosen to accomodate high pitched singing voices.  We use large versions of the pre-trained Wav2Vec 2.0 models found on the original authors' website \footnote{https://github.com/facebookresearch/fairseq/tree/main/examples/wav2vec}.  Our best performing model uses the ASR fine-tuned version of the Wav2Vec 2.0 model which was fine-tuned on 960 hours of Librispeech \cite{librispeech} data.  Similarly for HuBERT we experiment with the extra large version of the pre-trained model \footnote{https://github.com/facebookresearch/fairseq/tree/main/examples/hubert} that was trained on 60k hours of the Libri-Light dataset \cite{librilight}.  Both Wav2Vec 2.0 and HuBERT extract features at a rate of 50 Hz from 16 kHz audio.  Thus, we first downsample the audio and then upsample the extracted features to the same extraction frequency as the $f_0$ features (200 Hz).

We found that our $f_0$ encoders work well with $L=400$ bins and an embedding size of $F=256$.  Our PBTC architecture worked well with $K=10$ parallel layers.  We extract Wav2Vec 2.0 features from layer 12 of the transformer with dimensionality $D=1024$ and we extract HuBERT features from the penultimate layer of the transformer with dimensionality $D=1024$ as well.  Our speaker embedding module is a simple look up table with dimensionality $S=128$.  We train our models for 5 million steps on 8 NVIDIA A100 GPUs.  We use a learning rate of 0.0002 with hyperparameters $\lambda_{recon}=40, \lambda_{fm}=1$.

\subsection{Experimental Setup}
\label{sec:experimental setup}

We evaluated our singing voice conversion model by experimenting with different $f_0$ encoding schemes and \textit{pre-trained} self-supervised representations.  Additionally, we implemented a baseline based on disentanglement training to learn pitch and speaker-invariant self-supervised representations.  Since no publicly available baselines were available we implemented everything ourselves.  We found that a disentanglement task based on mutual information minimization \cite{kang2021robust} lead to training convergence, but in general we found disentanglement training to be very unstable.

\subsection{Results}
\label{sec: results}

We performed four different MOS studies to evaluate the various systems that we implemented\footnote{Audio samples: https://tkj516.github.io/Self-Supervised-Singing-Voice-Conversion/}.  If a system uses ASR fine-tuned self supervised features, we will mention it explicitly.   We synthesized singing samples in 4 speaker voices from our internal speech dataset, with 10 random singing segments from the dev sets. For each MOS study, we asked 300 raters to rate the quality/speaker similarity of 10 randomly chosen samples on a scale of 1-5.  All studies were performed using MTurk and unreliable raters were filtered out before computing the scores.  We tested the sytems along two major axis --- 1) \textit{audio quality}, i.e., how natural and human-like the audio clips sounded and 2) \textit{target singer/speaker similarity}, i.e., how similar did a voice converted audio clip sound to a reference sample from the same singer/speaker.
\vspace{-0.5em}
\subsubsection{MOS Study 1: Naturalness across $f_0$ encoding schemes and baselines.}

We fix the self-supervised representation to HuBERT.  We generate converted voice samples using our different $f_0$ encoding schemes.  A baseline based on disentanglement training and one trained only on singing voice samples is also included.  We also experiment with $f_0$ shifting during inference.  As shown in Table \ref{table:1}, an $f_0$ encoding method that uses the PBTC architecture has the highest audio quality.  We also see that training only on singing samples and with additional disentanglement losses leads to a drop in performance.
\vspace{-0.5em}
\subsubsection{MOS Study 2: Target singer similarity with fixed self-supervised features.}

We fix the self-supervised feature to HuBERT and provide the raters with a reference ground truth audio sample of the target singer and a converted sample from our model.  As shown in Table \ref{table:2}, $f_0$ shifting during inference significantly improves the speaker similarity.
\vspace{-0.5em}
\subsubsection{MOS Studies 3 \& 4: Varying the self-supervised features along with $f_0$ shifting during inference.}

We now vary the self-supervised representations being used and perform an $f_0$ shift to retain as much target speaker identity as possible.  As shown in Table \ref{table:3}, using ASR fine-tuned Wav2Vec 2.0 features along with a PBTC encoder results in excellent audio quality.  It significantly outperforms baselines that use vanilla Wav2Vec 2.0 features without ASR fine-tuning. 

We also test for target singer identity similarity as shown in Table~\ref{table:4}.  The results show similar trends with the architectures using ASR fine-tuned features retaining the target  singer identity well.  Surprisingly, the architecture that uses HuBERT + PBTC also achieves higher target singer similarity, probably suggesting that the HuBERT embedding encodes less speaker dependent information in it.

\section{Conclusion}

In this paper we propose a singing voice conversion architecture that uses ASR fine-tuned Wav2Vec 2.0 features along with a specialized $f_0$ encoder.  Experiments show that an $f_0$ encoder based on a parallel bank of transposed convolutions (PBTC) leads to the best audio quality and that when combined with $f_0$ distribution matching during inference helps retain target singer identity.  We conduct extensive MOS studies to test our architecture and detail how our model is implemented.  Additionally, since our model was trained on both speech and singing data, it can be used to convert a song to the voice of target speaker for which no singing data is available.  

For future work, we are looking to use disentangled ASR fine-tuned Wav2Vec 2.0 features for speech to singing voice conversion, with the main challenge being the unsupervised learning of an alignment between speech and melody features.  We are also looking into additional architectures for singing voice synthesis with quantized self-supervised representations.

\begin{table}[!t]
\centering
\begin{tabular}{l|l}
\hline
\multicolumn{1}{c|}{\textbf{System}} & \multicolumn{1}{c}{\textbf{MOS}} \\ \hline
Ground Truth                          & $4.22 \pm 0.04$                       \\ \hline
Q-LUT (Sing only)                     & $3.44 \pm 0.07$                       \\ \hline
Q-LUT                                 & $3.69 \pm 0.04$                       \\ \hline
Q-LUT (Disentanglement)               & $3.37 \pm 0.07$                       \\ \hline
\textbf{PBTC}                                  & $\mathbf{3.76 \pm 0.06}$                       \\ \hline
Q-LUT + f0 shift                      & $3.66 \pm 0.06$                       \\ \hline
PBTC + f0 shift                       & $3.68 \pm 0.06$                       \\ \hline
\end{tabular}
\caption{MOS Study 1: Comparison of audio quality with different singing voice conversion models while using HuBERT self-supervised features.\vspace{-0.5em}}
\label{table:1}
\end{table}

\begin{table}[!t]
\centering
\begin{tabular}{l|l}
\hline
\multicolumn{1}{c|}{\textbf{System}} & \multicolumn{1}{c}{\textbf{MOS}} \\ \hline
Q-LUT                                 & $3.70 \pm 0.04$                       \\ \hline
Q-LUT (Disentanglement)               & $3.57 \pm 0.04$                       \\ \hline
\textbf{Q-LUT + f0 shift}                      & $\mathbf{3.79 \pm 0.04} $                      \\ \hline
\textbf{PBTC + f0 shift}                       & $\mathbf{3.80 \pm 0.04}$                       \\ \hline
\end{tabular}
\caption{MOS Study 2: Comparison of target singer/speaker similarity with different singing voice conversion models while using HuBERT self-supervised features.\vspace{-0.5em}}
\label{table:2}
\end{table}

\begin{table}[!t]
\centering
\begin{tabular}{l|l}
\hline
\multicolumn{1}{c|}{\textbf{System}} & \multicolumn{1}{c}{\textbf{MOS}} \\ \hline
Wav2Vec2 + f0 shift (Q-LUT)           & $3.67 \pm 0.05$                       \\ \hline
Wav2Vec2 + f0 shift (PBTC)            & $3.77 \pm 0.05$                       \\ \hline
HuBERT + f0 shift (Q-LUT)             & $3.81 \pm 0.04$                       \\ \hline
HuBERT + f0 shift (PBTC)              & $3.81 \pm 0.04$                       \\ \hline
\textbf{Wav2Vec2-ASR + f0 shift (Q-LUT)}       & $\mathbf{3.83 \pm 0.06}$                       \\ \hline
\textbf{Wav2Vec2-ASR + f0 shift (PBTC)}        & $\mathbf{3.91 \pm 0.05}$              \\ \hline
\end{tabular}
\caption{MOS Study 3: Comparison of audio quality while varying the self-supervised feature and $f_0$ encoder.\vspace{-0.5em}}
\label{table:3}
\end{table}

\begin{table}[!t]
\centering
\begin{tabular}{l|l}
\hline
\multicolumn{1}{c|}{\textbf{System}} & \multicolumn{1}{c}{\textbf{MOS}} \\ \hline
Wav2Vec2 + f0 shift (Q-LUT)           & $3.79 \pm 0.04$                       \\ \hline
Wav2Vec2 + f0 shift (PBTC)            & $3.75 \pm 0.05$                       \\ \hline
HuBERT + f0 shift (Q-LUT)             & $3.82 \pm 0.04$                       \\ \hline
\textbf{HuBERT + f0 shift (PBTC)}              & $\mathbf{3.85 \pm 0.06}$                       \\ \hline
Wav2Vec2-ASR + f0 shift (Q-LUT)       & $3.84 \pm 0.05$                       \\ \hline
Wav2Vec2-ASR + f0 shift (PBTC)       & $3.83 \pm 0.06$                       \\ \hline
\end{tabular}
\caption{MOS Study 4: Comparison of target singer/speaker similarity while varying the self-supervised feature and $f_0$ encoder.\vspace{-1em}}
\label{table:4}
\end{table}

\section{Acknowledgements}

We would like to thank Yossi Adi for insightful discussions and feedback.

\vfill\pagebreak


\bibliographystyle{IEEEbib}
\bibliography{strings}

\begin{thebibliography}{10}

\bibitem{saino2006hmm}
Keijiro Saino, Heiga Zen, Yoshihiko Nankaku, Akinobu Lee, and Keiichi Tokuda,
\newblock ``An hmm-based singing voice synthesis system,''
\newblock in {\em Ninth International Conference on Spoken Language
  Processing}, 2006.

\bibitem{oura2010recent}
Keiichiro Oura, Ayami Mase, Tomohiko Yamada, Satoru Muto, Yoshihiko Nankaku,
  and Keiichi Tokuda,
\newblock ``Recent development of the hmm-based singing voice synthesis
  system—sinsy,''
\newblock in {\em Seventh ISCA Workshop on Speech Synthesis}, 2010.

\bibitem{nakamura2014hmm}
Kazuhiro Nakamura, Keiichiro Oura, Yoshihiko Nankaku, and Keiichi Tokuda,
\newblock ``Hmm-based singing voice synthesis and its application to japanese
  and english,''
\newblock in {\em 2014 IEEE International Conference on Acoustics, Speech and
  Signal Processing (ICASSP)}. IEEE, 2014, pp. 265--269.

\bibitem{kobayashi2014statistical}
Kazuhiro Kobayashi, Tomoki Toda, Graham Neubig, Sakriani Sakti, and Satoshi
  Nakamura,
\newblock ``Statistical singing voice conversion with direct waveform
  modification based on the spectrum differential,''
\newblock in {\em Fifteenth Annual Conference of the International Speech
  Communication Association}. Citeseer, 2014.

\bibitem{kobayashi2015statistical}
Kazuhiro Kobayashi, Tomoki Toda, Graham Neubig, Sakriani Sakti, and Satoshi
  Nakamura,
\newblock ``Statistical singing voice conversion based on direct waveform
  modification with global variance,''
\newblock in {\em Sixteenth Annual Conference of the International Speech
  Communication Association}. Citeseer, 2015.

\bibitem{oord2016wavenet}
Aaron van~den Oord, Sander Dieleman, Heiga Zen, Karen Simonyan, Oriol Vinyals,
  Alex Graves, Nal Kalchbrenner, Andrew Senior, and Koray Kavukcuoglu,
\newblock ``Wavenet: A generative model for raw audio,''
\newblock {\em arXiv preprint arXiv:1609.03499}, 2016.

\bibitem{kalchbrenner2018efficient}
Nal Kalchbrenner, Erich Elsen, Karen Simonyan, Seb Noury, Norman Casagrande,
  Edward Lockhart, Florian Stimberg, Aaron Oord, Sander Dieleman, and Koray
  Kavukcuoglu,
\newblock ``Efficient neural audio synthesis,''
\newblock in {\em International Conference on Machine Learning}. PMLR, 2018,
  pp. 2410--2419.

\bibitem{kong2020hifi}
Jungil Kong, Jaehyeon Kim, and Jaekyoung Bae,
\newblock ``Hifi-gan: Generative adversarial networks for efficient and high
  fidelity speech synthesis,''
\newblock {\em Advances in Neural Information Processing Systems}, vol. 33, pp.
  17022--17033, 2020.

\bibitem{nachmani2019unsupervised}
Eliya Nachmani and Lior Wolf,
\newblock ``Unsupervised singing voice conversion,''
\newblock {\em arXiv preprint arXiv:1904.06590}, 2019.

\bibitem{deng2020pitchnet}
Chengqi Deng, Chengzhu Yu, Heng Lu, Chao Weng, and Dong Yu,
\newblock ``Pitchnet: Unsupervised singing voice conversion with pitch
  adversarial network,''
\newblock in {\em ICASSP 2020-2020 IEEE International Conference on Acoustics,
  Speech and Signal Processing (ICASSP)}. IEEE, 2020, pp. 7749--7753.

\bibitem{polyak2020unsupervised}
Adam Polyak, Lior Wolf, Yossi Adi, and Yaniv Taigman,
\newblock ``Unsupervised cross-domain singing voice conversion,''
\newblock {\em arXiv preprint arXiv:2008.02830}, 2020.

\bibitem{wang2021towards}
Chao Wang, Zhonghao Li, Benlai Tang, Xiang Yin, Yuan Wan, Yibiao Yu, and Zejun
  Ma,
\newblock ``Towards high-fidelity singing voice conversion with acoustic
  reference and contrastive predictive coding,''
\newblock {\em arXiv preprint arXiv:2110.04754}, 2021.

\bibitem{hsu2021hubert}
Wei-Ning Hsu, Benjamin Bolte, Yao-Hung~Hubert Tsai, Kushal Lakhotia, Ruslan
  Salakhutdinov, and Abdelrahman Mohamed,
\newblock ``Hubert: Self-supervised speech representation learning by masked
  prediction of hidden units,''
\newblock {\em IEEE/ACM Transactions on Audio, Speech, and Language
  Processing}, vol. 29, pp. 3451--3460, 2021.

\bibitem{polyak2021speech}
Adam Polyak, Yossi Adi, Jade Copet, Eugene Kharitonov, Kushal Lakhotia,
  Wei-Ning Hsu, Abdelrahman Mohamed, and Emmanuel Dupoux,
\newblock ``Speech resynthesis from discrete disentangled self-supervised
  representations,''
\newblock {\em arXiv preprint arXiv:2104.00355}, 2021.

\bibitem{baevski2020wav2vec}
Alexei Baevski, Yuhao Zhou, Abdelrahman Mohamed, and Michael Auli,
\newblock ``wav2vec 2.0: A framework for self-supervised learning of speech
  representations,''
\newblock {\em Advances in Neural Information Processing Systems}, vol. 33, pp.
  12449--12460, 2020.

\bibitem{devlin2018bert}
Jacob Devlin, Ming-Wei Chang, Kenton Lee, and Kristina Toutanova,
\newblock ``Bert: Pre-training of deep bidirectional transformers for language
  understanding,''
\newblock {\em arXiv preprint arXiv:1810.04805}, 2018.

\bibitem{graves2006connectionist}
Alex Graves, Santiago Fern{\'a}ndez, Faustino Gomez, and J{\"u}rgen
  Schmidhuber,
\newblock ``Connectionist temporal classification: labelling unsegmented
  sequence data with recurrent neural networks,''
\newblock in {\em Proceedings of the 23rd international conference on Machine
  learning}, 2006, pp. 369--376.

\bibitem{webber2020hider}
Jacob Webber, Olivier Perrotin, and Simon King,
\newblock ``Hider-finder-combiner: an adversarial architecture for general
  speech signal modification,''
\newblock in {\em Interspeech 2020-21st Annual Conference of the International
  Speech Communication Association}, 2020.

\bibitem{duan2013nus}
Zhiyan Duan, Haotian Fang, Bo~Li, Khe~Chai Sim, and Ye~Wang,
\newblock ``The nus sung and spoken lyrics corpus: A quantitative comparison of
  singing and speech,''
\newblock in {\em 2013 Asia-Pacific Signal and Information Processing
  Association Annual Summit and Conference}. IEEE, 2013, pp. 1--9.

\bibitem{choi2020children}
Soonbeom Choi, Wonil Kim, Saebyul Park, Sangeon Yong, and Juhan Nam,
\newblock ``Children’s song dataset for singing voice research,''
\newblock in {\em International Society for Music Information Retrieval
  Conference (ISMIR)}, 2020.

\bibitem{kim2018crepe}
Jong~Wook Kim, Justin Salamon, Peter Li, and Juan~Pablo Bello,
\newblock ``Crepe: A convolutional representation for pitch estimation,''
\newblock in {\em 2018 IEEE International Conference on Acoustics, Speech and
  Signal Processing (ICASSP)}. IEEE, 2018, pp. 161--165.

\bibitem{librispeech}
Vassil Panayotov, Guoguo Chen, Daniel Povey, and Sanjeev Khudanpur,
\newblock ``Librispeech: An asr corpus based on public domain audio books,''
\newblock in {\em 2015 IEEE International Conference on Acoustics, Speech and
  Signal Processing (ICASSP)}, 2015, pp. 5206--5210.

\bibitem{librilight}
J.~Kahn, M.~Riviere, W.~Zheng, E.~Kharitonov, Q.~Xu, P.E. Mazare, J.~Karadayi,
  V.~Liptchinsky, R.~Collobert, C.~Fuegen, T.~Likhomanenko, G.~Synnaeve,
  A.~Joulin, A.~Mohamed, and E.~Dupoux,
\newblock ``Libri-light: A benchmark for {ASR} with limited or no
  supervision,''
\newblock in {\em {ICASSP} 2020 - 2020 {IEEE} International Conference on
  Acoustics, Speech and Signal Processing ({ICASSP})}. may 2020, {IEEE}.

\bibitem{kang2021robust}
Woo~Hyun Kang, Jahangir Alam, and Abderrahim Fathan,
\newblock ``Robust speech representation learning via flow-based embedding
  regularization,''
\newblock {\em arXiv preprint arXiv:2112.03454}, 2021.

\end{thebibliography}

\end{document}